\documentclass[aps,amssymb,amsmath, twocolumn, pra, superscriptaddress]{revtex4}
\usepackage{graphicx}
\usepackage{epsfig}
\usepackage{dcolumn}
\usepackage[up]{subfigure}
\usepackage{float}
\usepackage{dsfont}	
\usepackage{relsize}
\usepackage{textcomp}

\usepackage{epstopdf}

\usepackage{float}
\usepackage{ragged2e}
\usepackage{braket}
\usepackage{bbold}
\usepackage{bbm}

\usepackage[font=small,labelfont=bf,justification=raggedright, format=plain]{caption}
\usepackage{setspace}
\usepackage[colorlinks, citecolor = magenta]{hyperref}
\usepackage{cancel}
\usepackage{comment}  
\makeatletter
\begin{document}
\title{Quantum illumination using polarization-path entangled single photons for low reflectivity object detection in noisy background}
\author{K. Muhammed Shafi}
\affiliation{Quantum Optics \& Quantum Information, Department of Instrumentation and Applied Physics, Indian Institute of Science, Bengaluru 560012, India}
\author{A. Padhye}
\affiliation{Quantum Optics \& Quantum Information, Department of Instrumentation and Applied Physics, Indian Institute of Science, Bengaluru 560012, India}
\author{C. M. Chandrashekar}
\email{chandracm@iisc.ac.in}
\affiliation{Quantum Optics \& Quantum Information, Department of Instrumentation and Applied Physics, Indian Institute of Science, Bengaluru 560012, India}
\affiliation{The Institute of Mathematical Sciences, C. I. T. Campus, Taramani, Chennai 600113, India}
\affiliation{Homi Bhabha National Institute, Training School Complex, Anushakti Nagar, Mumbai 400094, India}

\begin{abstract}
Detecting object with low reflectivity embedded within a noisy background is a challenging task. Quantum correlations between pairs of quantum states of light, though are highly sensitive to background noise and losses, offer advantages over traditional illumination methods.  Instead of using correlated photon pairs which are sensitive,  we experimentally demonstrate the advantage of using heralded single-photons entangled in polarization and path degree of freedom for quantum illumination. In the study, the object of different reflectivity is placed along the path of the signal in a variable thermal background before taking the joint measurements and calculating the quantum correlations.  We show the significant advantage of using non-interferometric measurements along the multiple paths for single photon to isolate the signal from the background noise and outperform in detecting and ranging the low reflectivity objects even when the signal-to-noise ratio is as low as 0.03. Decrease in visibility of polarization along the signal path also results in similar observations. This will have direct relevance to the development of single-photon based quantum LiDAR and quantum imaging.
\end{abstract}

\maketitle


\section{\label{sec1}Introduction}

Quantum correlations in the form of entanglement is a salient feature of quantum mechanics and is central to many quantum information processing protocols\,\cite{FSS19, PHF16, JSW2000, BPM97, PBG18}. However,  they are  highly sensitive to environmental noise and can be easily destroyed affecting advantages gained by such nonclassical correlations.  Quantum illumination (QI) which uses quantum correlations between pair of photons for object detection in a noisy environment is an exception\,\cite{SL08, TEG08, LBD13, UR09}.  Known approaches for QI rely upon two entangled pair of beams in the form of signal and idler as probe for object detection, wherein signal beam is sent to a region of space containing object merged in background noise and the idler  beam is stored locally until the signal reflects from the object.  The enhancement of performance of QI over classical analog is made possible by using detection and joint measurement techniques which capture the nonclassical correlations between the stored idler and the reflected signal by isolating background noise.  QI measurements primarily focus on reducing uncertainty in unknown parameter estimation using quantum correlation. Thus, QI extends principles of target detection accuracy, ranging sensitivity, and degree of resilience towards preponderant noise from conventional radar technology to quantum metrology\,\cite{PVS20, MR20}. 

The general formalism for quantum sensing originates from quantum channel discrimination model employed for target detection in thermal background. The model is based on pioneering work by Helstrom in 1976\,\cite{CWH76} on quantum hypothesis testing for minimum error probability which discriminates between two channels- one with the input state reflected from the target and other with the thermal noise implying presence or absence of the target, respectively\,\cite{PLL19}. It further led to counter-intuitive observations reported by Sacchi in 2005 that entangled input states enhance the discrimination of two entanglement-breaking channels with minimal error probability\,\cite{MFS71,MFS72}. In 2008 Lloyd translated these concepts and proposed the first theoretical framework for QI using entangled photons being sent repeatedly to detect a weakly reflecting object immersed in noisy background\,\cite{SL08}. He showed that the entangled probe state reduces the number of trials needed to detect the object by a factor of number of modes per detection event, even when signal to noise ratio (SNR)  $< 1$. Further, a more general model was considered to include multi-photon Hilbert space\,\cite{JSL09}. Soon after,  continuous variable based QI protocol was reported\,\cite{TEG08}. It showed 6 dB gain in error probability exponent using computational tools\,\cite{ACM07,PL08,CMM08} for a Gaussian probe state over a coherent state system. In order to realize this improvement using Gaussian state, two optical receivers were proposed viz. an optical parametric amplifier with small gain and a phase conjugate receiver with balanced detection\,\cite{GIE09}. Both showed 3 dB error-exponent gain achievable through practical QI protocol.  Recent theoretical studies have also reported the improved efficiency of QI when hyperentangled probe states are used\,\cite{PSC21, KC22}.

Even though several theoretical QI protocols were proposed, their experimental realization has been challenging task. Mainly due to the unavailability of quantum optimal receivers which involves the difficulty in devising perfect mode-matching for joint phase-sensitive measurements between the reflected signal and the stored reference beams. The first experimental demonstration of QI was based on phase-insensitive intensity measurements\,\cite{LBD13, LBO14}.  They showed improvement in SNR for target detection using photon-number correlations between twin beams from spontaneous parametric down conversion (SPDC),  wherein object and thermal noise was introduced in one of the beams. Generalised Cauchy-Schwarz parameter was used to quantify nonclacicality of the measured output and nonclassical signal was recorded upto SNR $\approx 0.5$ when object reflectivity was $50\%$. QI protocols have also been explored in the microwave regime in which a target was probed by microwave wavelength and detected at optical frequency by using electro-optomechanical converters\,\cite{BGW14}.  It was followed by another report where they used an optical parametric amplifier quantum receiver as proposed in theory but of semi-optimal nature and compared its performance with an optimal classical illumination system\,\cite{ZMW15}. A QI scheme was also reported to show 10 times improvement in SNR over its classical analogue, even without making a joint measurement on the signal and reference beams\,\cite{CVB19}. In another QI implementation by using a maximally entangled state as the optimum probe state to illuminate the potential target, surpassing the classical limit for up to 40\%, while approaching the quantum limit imposed by the Helstrom limit\,\cite{FL2021} was reported. In the last few years, there has been significant development in using temporal, spectral, and polarization correlation for target detection from noisy backgrounds\,\cite{YB2020, HK2020}.  Experimental demonstration of QI of diffusively reflecting object where $SNR > 40dB$ has also been reported\,\cite{EBS19}. 

Apart from several quantum metrology protocols, QI methods are used for other emerging applications which includes quantum communication. Following the theoretical proposal by Shapiro\,\cite{JHS09}, experimental implementation were also reported showing the immunity towards passive eavesdropping\,\cite{ZTZ13, SZW14},  Alice's error probability was shown to be much lower than that of Eve's even though Bob's amplifier destroyed the entanglement. QI systems have also been extended further to quantum imaging applications \cite{MLS17, GMT20, ZYW21, GM2021, ZLU22}.
		
Single photon entangled in internal degree of freedom like path and polarization provide a natural representation of quantum bits and are likely to play an important role in the future development of quantum technologies\,\cite{SS17, Thew2021, Thew2022, Anto2022}. Here in a first of its kind, we experimentally demonstrate the advantage of using single photons entangled in polarization and path degree of freedom for QI.  We employ heralded single photons from the photon pairs generated using SPDC process, one of the photon from the pair is retained as an idler  photon and used for heralding,  whereas the other \textbf{signal} photon is entangled in the polarization and path degree of freedom and used along two paths in the experimental setup. Unlike earlier protocols know for QI, we have used photon pairs generated from SPDC process and not the entangled photon pairs. In the scheme, three pathways are employed for the two photons.  One  of the pathways is used for heralding the polarization-path entangled photon and the other two pathways are used as a signal and reference paths which are sent towards the object and directly to the receiving unit, respectively.  An object of different reflectivity $\eta$ is placed along the path of the signal and the controlled noise in the form of thermal background is introduced along the path of the signal before taking joint measurements and calculating the quantum correlations. Bell's inequality violation in the form of Clauser, Horne, Shimony, and Holt (CHSH) parameter $S > 2$ is employed to quantify quantum correlation and detect the presence or absence of object in presence of background noise. 
\begin{figure}[h!]
\centering
\hskip 0.6 in \includegraphics[width=0.5\textwidth]{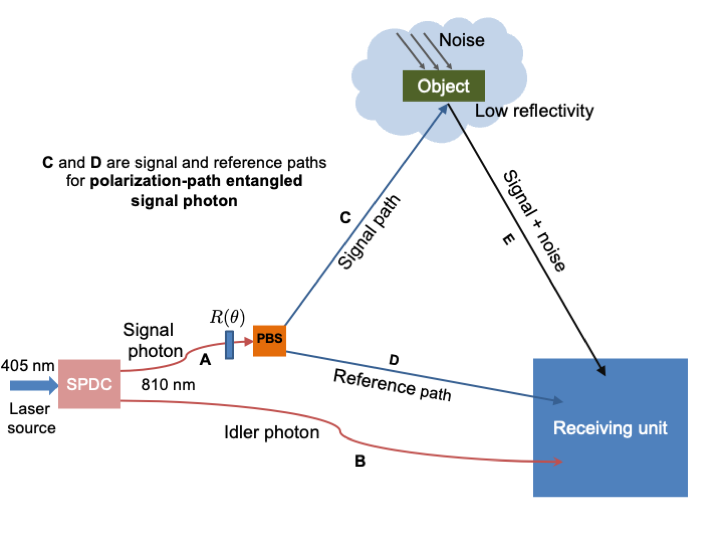}
\caption{Schematic of the quantum illumination protocol using polarization-path entangled single photons.  Spontaneous parametric down conversion process is used to generate photons pairs and one of them, signal photon along path {\bf A} is entangled in polarization and path degree of freedom and sent along signal path, {\bf C}  and reference path, {\bf D}.  Another photon from the pair along path {\bf B} is used as idler for heralding signal  photons.  An object of different reflectivity is placed in the signal path with variable background noise. Entanglement between polarization and path degree of freedom using coincidence counts of photons along paths paths {\bf D} and {\bf E} is calculated with photons along {\bf B}.  Bell's inequality violation in the form of CHSH parameter $S$ is used to quantify quantum correlation and object detection.}
\label{QI-sc}
\end{figure}

 In Fig.\,\ref{QI-sc} the schematic of the protocol for QI using polarization-path entangled single photons at 810 nm from SPDC process is shown.   From the pair of down converted photon, signal photon is entangled in polarization and path degree of freedom and the idler photon is used as a reference photon for heralding.  One of the two paths of the polarization-path entangled single photon  is send towards the object (signal path) and other path is used as a reference path.  Using the coincidence detection of photons along the three paths, CHSH parameter is calculated to quantify quantum correlation.  To calculate CHSH parameter using coincidence counts along multiple paths, we have used both, interferometric and non-interferometric approach  at the receiving unit.  We show that both the approaches work well when we have object of different reflectivity in the path of the signal and in absence of background noise.  In presence of background noise, a significant advantage is seen when only non-interferometric measurements are used for detecting the object.  The demonstrated protocol using non-interferometric approach  isolates the background thermal noise from the signal and help in detecting the object by returning quantum correlation value of $S>2$ even when signal is buried under the noise with SNR as low as 0.03.  Even when we can't record quantum correlation $S<2$, for a range $ 1.44 <   S <  2$ showing the classical correlation as residual of quantum correlation can still be used to detect the low reflectivity object immersed in noisy background with SNR as low as 0.03 corresponding to -15 dB.   The use of CHSH parameter compared to earlier reported result using  Cauch-Schwarz parameter to measure nonclasscaility\,\cite{LBD13} upto SNR$\approx 0.5$ and the use of three path approach for two photons  has resulting in significant suppression of background noise in our scheme. The  effect of decrease in polarization visibility which models the polarization scattering by the object has also been studied and even when the polarization visibility is as low are 0.2 we can identify the presence of object.   The result reported here using heralded single photons of 810 nm wavelength will have direct relevance to development of quantum LiDAR and can be adopted to other wavelengths and on demand single photon sources.

\section{Polarzation-path entangled single photons for QI} 
\label{sec2} 


The state of the single photon in equal superposition of two linearly polarized states, $|h\rangle$ and $|v\rangle$  when passed through the polarizing beam splitter (PBS) can be written in the form,
\begin{align}
|\Psi_0 \rangle =  \frac{1}{\sqrt{2}} \left [ |h\rangle |0\rangle -  |v\rangle |1\rangle \right ].
\label{intialsta}
\end{align}
The states $|0\rangle$ and $|1\rangle$ are the two polarization dependent paths for the photons and we will refer to them as the reference path and signal path, respectively. 
 The preceding  state is maximally entangled in polarization and path degree of freedom.   When the photon passes through the object with reflectivity $ 0 \leq  \eta \leq 1$ along the signal path, the effect of the object on the photons state can be written in the form of a controlled operator causing loss along the path of the signal,
 \begin{align}
T(\eta )  =   |h\rangle \langle h | \otimes   | 0 \rangle \langle 0 |  + \sqrt{\eta}  | v \rangle \langle v | \otimes   | 1 \rangle \langle 1 | .
\end{align}

The density matrix of the polariztion-path entangled photon at the receiving end of the signal and reference path will be in the form,
\begin{align}
\rho (\eta) = T (\eta)\left  (  |\Psi_0 \rangle \langle \Psi _0 | \right ) T (\eta)^{\dagger}.
\end{align}
Below we present two measurement procedure to analyze  the photons arriving along both, signal and reference paths of identical path lengths. 
 \begin{figure}[h!]
\centering
\includegraphics[width=0.49\textwidth]{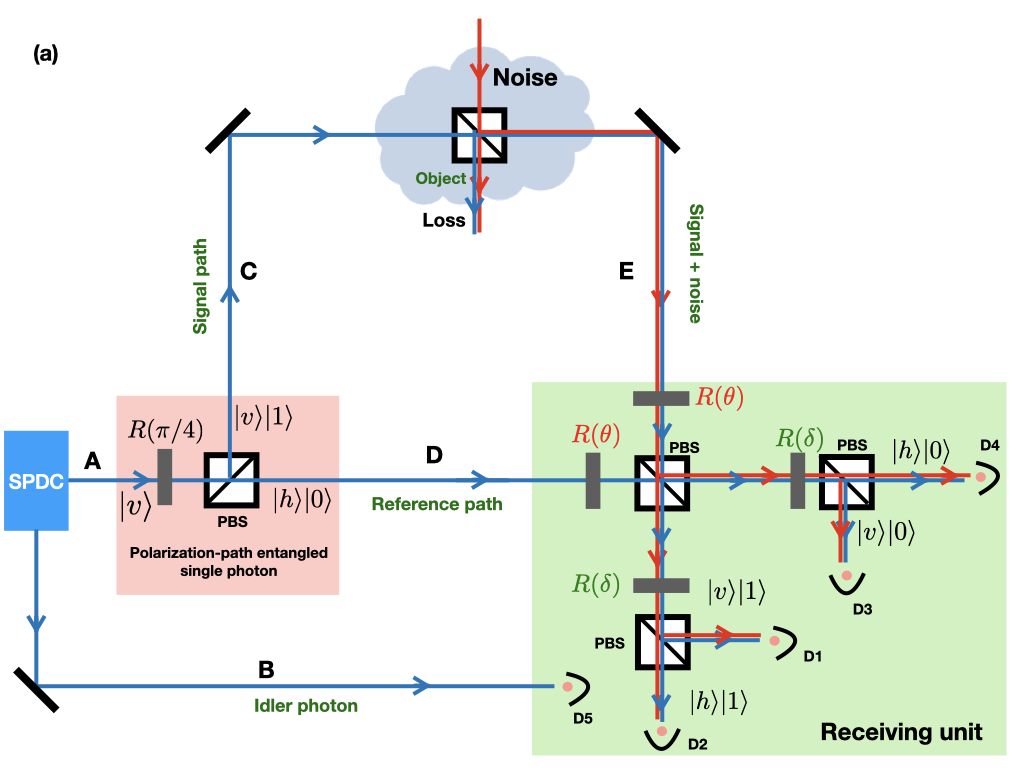}
\includegraphics[width=0.49\textwidth]{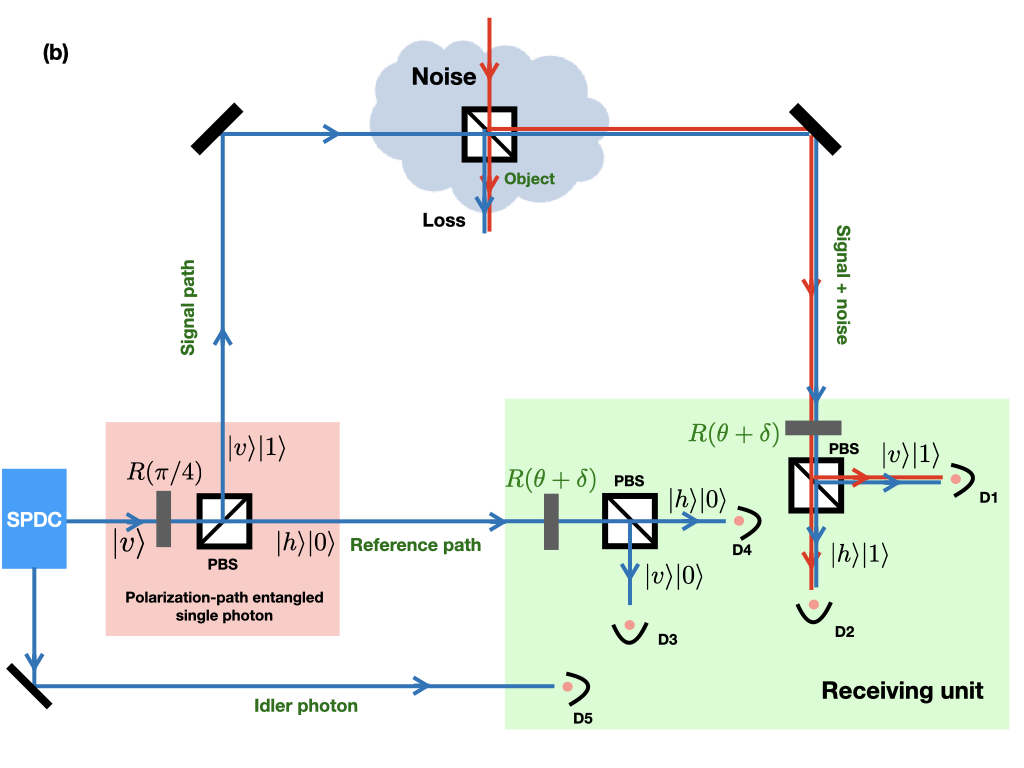}
\caption{Schematic of the experimental setup  for QI using polarization-path entangled single-photon. The heralded single-photon entangled state are sent across two paths where one path acts as reference and reaches the receiving unit directly and the other path intercepts with the object and reflects back to the receiver.  Two different measurement procedures are shown to calculate quantum correlation. (a) An interferometric approach where photons from both the paths are made to interfere.  (b) A non-interferometric approach where photons from both the paths do not interference.  PBS and polarization rotator $R(.)$  in the scheme are used to control the splitting ratio along the paths by varying $\theta$ and $\delta$.  Using the coincidence counts of photons along different paths with the idler photons for different combination of  $( \theta, \delta)$, and $(\theta^{\prime}, \delta^{\prime})$ we can calculate  CHSH parameter, $S$.  When the object reflectivity, $\eta = 1$ and in absence of background noise, both the procedure are equivalent and give same CHSH parameter.  }
\label{QI-inter}
\end{figure}

Ideal method  to calculate quantum correlations from single photons in path degree of freedom would involve interference of the two paths and probabilities  of output states.  To quantify quantum correlation, we will calculate CHSH parameter $S$,
\begin{align}
S = \lvert  E(\theta, \delta) - E(\theta, \delta^{\prime})   +   E(\theta^{\prime}, \delta) + E(\theta^{\prime}, \delta^{\prime}  ) \rvert
\end{align} 
 where $E(\theta , \delta) = P_{h0} (\theta, \delta) + P_{v1}(\theta, \delta) - P_{h1}(\theta, \delta) - P_{v0}(\theta, \delta)$.  The $P_{ij}$'s are the probabilities of different basis states of the photon in polarization and path composition.  The parameter $\theta$ and $\delta$ are the angles that  that represent the polarization of photons.  Different basis states of the photon in polarization and path degree of freedom can be obtained  using the combination of the polarization rotator $R(\theta)  (R(\delta))$  and polarizing beam splitter (PBS) along the paths of the photons.  In Fig.\,\ref{QI-inter}(a) the schematic of the combination of  polarization rotator  $R(\cdot)$  and PBS along the interfering paths of the photons is shown.  The effect of the polarization rotator along both the paths before they interfere at PBS and after they interfere on the state of polarization-path entangled photon can be written in the form, 
\begin{align}
&\rho(\theta, \delta )_{I} =    \left ( \mathbbm{1} \otimes R(\delta) \right )( R(\theta) \otimes \mathbbm{1} )  \rho(\eta) (R(\theta) \otimes \mathbbm{1} )^{\dagger}( \mathbbm{1} \otimes R(\delta) )^{\dagger}
\end{align}
where  polarization rotator, $R(\theta) ~ (R(\delta))$  are given by the form,
\begin{align}
R(\theta) = \begin{bmatrix}
	~\cos(\theta) & -\sin(\theta) \\
	~\sin(\theta) & ~~\cos(\theta)
	\end{bmatrix}.
\end{align}	
It can be practically realized using the  half-wave plate (HWP)  rotated by an angle $\kappa/2$, $H(\kappa) = \begin{bmatrix}
	~\cos(2 \kappa) & ~~\sin(2\kappa) \\
	~\sin(2\kappa) & -\cos(2\kappa)
	\end{bmatrix}$  where $R(\theta) \equiv H (\kappa/2)$. The probabilities,  $P_{h0} = \rho(\theta, \delta )_{44}$,  $P_{v1} = \rho(\theta, \delta )_{11}$, $P_{h1} = \rho(\theta, \delta )_{22}$ and $P_{v0} = \rho(\theta, \delta )_{33}$ are the the diagonal elements of the density matrix.  However,  coherence of single photon along multiple paths is extremely hard to achieve experimentally for longer  path lengths. Therefore, an equivalent non-interferometric method can be effectively used to calculate the quantum correlations in single  photon states\,\cite{SGP21}.  The equivalent density matrix and the probabilities of the basis state of the composite system can be obtained by performing an identical rotation independently along both the paths. The schematic of the combination of polarization rotator and PBS along both the non-interfering paths are shown in Fig.\,\ref{QI-inter}(b) and the state can be written in the form,  
	
\begin{align}
\rho(\theta, \delta )_{NI} =  ( R(\theta + \delta) \otimes \mathbbm{1} )  \rho(\eta)  ( R(\theta + \delta) \otimes \mathbbm{1} )^{\dagger}.
\end{align}

For combinations of $\theta$ and $\delta$ we can record the maximum value of CHSH parameter, $S = 2\sqrt{2}$ when object reflectivity, $\eta = 1$ for both, interferometric and non-interferometric approach. 
 \begin{figure}[h!]
\centering
\includegraphics[width=0.49\textwidth]{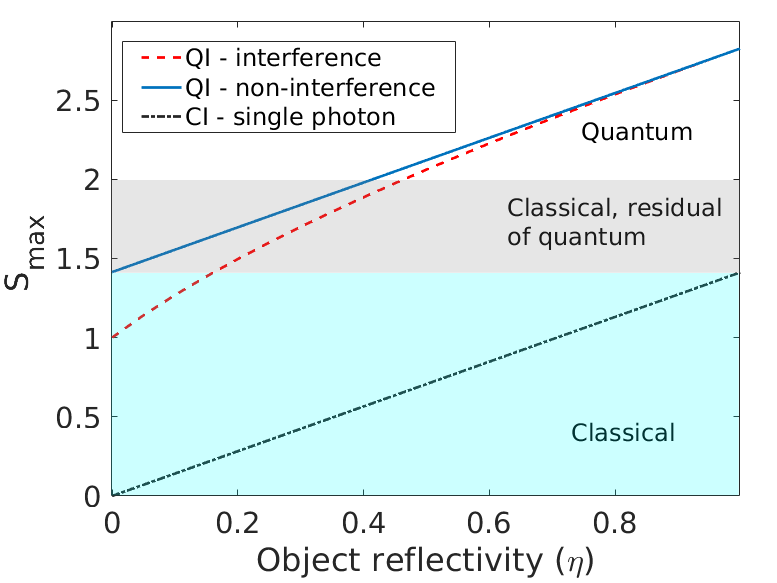}
\caption{Theoretically calculated CHSH parameter $S_{max}$ when object of different reflectivity $\eta$ is placed along the signal path using polarization-path entangled single photons as probe and different measurement schemes at the receiving unit. Non-interference approach shows advantage at low reflectivity regime. To show the classical regime, $S_{max}$ when single photons in superposition state is sent only across signal path is shown and the value $S_{max} < 1.44$. This allows us to use the classical correlation between polarization and path degree of freedom in the range $2 > S > 1.44$ as residual of quantum correlation to identify the object of very low reflectivity using non-interferometric measurement scheme.}
\label{QI-interNon}
\end{figure}
In Fig.\,\ref{QI-interNon}, $S_{max}$ as function of  $\eta$ is shown for both, interferometric and non-interferometric measurement schemes. One set of parameters we get maximum value when the initial state of the form given in Eq.\,\eqref{intialsta} and sent across signal and reference paths are $\theta = 0$, $\delta = \pi/16$, $\theta^{\prime} = 3\pi/16$ and $\delta^{\prime} = 5\pi/16$.  
We can note that the non-interference approach provides advantage at low reflectivity region by returning higher $S$ value.   This can be attribute the difference in value of $E(\theta, \delta)$, a straight forward calculation of combination of probabilities of finding photons along four paths in both, the interference and non-interference scheme. Only two of the probabilities varies with varying reflectivity in non-interference scheme and all the four varies in interference scheme affecting the overall value of $S$. In this work we only see it has a metric with difference in value of $S$ at low reflectivity regime. 

The same scheme can be turned to a classical illumination scheme by replacing the initial polarization-path entangled single photon state with the single photon in state $|\psi\rangle = \frac{1}{\sqrt{2}} (|h\rangle - |v\rangle )$ only along the signal path. For the choice of parameters given above, the measuring unit will not record any correlation in polarization and path degree of freedom. In the Fig.\,\ref{QI-interNon} we have show the value of $S_{max} < 1.44$ when measurement configuration as shown in Fig.\ref{QI-inter}(b) is used. This gives the upper bound on the value of maximum value of $S$ when single photon without being in correlation with its internal degree of freedom is used.  Therefore, even when $S < 2$, in absence of violation of Bell's inequality, we can use the classical correlation  $1.44 < S < 2 $ as the residual of the quantum correlation to identify the presence of object with low reflectivity. Since we don't observe the value of $S > 1.44$ when source is not in correlated state, the residual of quantum correlation,  $1.44 < S < 2$  can be attributed to the presence of correlation in the initial state of photons source used for illumination. We also want to make a note that the mathematical framework of single photon in signal and reference path guided in to the receiving unit is in the form of two step discrete-time quantum walk where the HWP and PBS are quantum coin operation and polarization dependent shift operators. Thus, various other configuration of parameters can be adopted to control and measure correlation between the polarization and path degree of freedom in presence of noise.

\noindent
{\it Thermal and depolarizing noise:} When thermal noise is introduced in the form of white light along the path of signal, the noisy photons also get detected along with the photons from the signal path but their random polarization will only result in the increase in offset of the photons detected in the detectors and ideally only the photons from signal will contribute to change in the polarization when rotated using the HWP. Therefore, until the fluctuation in thermal noise supersedes the change in signal photon counts in detectors with HWP, we will be able to get a reliable $S$ value and help in detected the presence of object. However, in interferometric measurement scheme, since all the four detectors receive noisy photons, they can contribute to false coincidence counts resulting in decrease in $S$ parameter. In non-interferometric scheme, the reference path which does not receive any noisy photons will reduce the false coincidence counts contributing for its robustness against noisy photons. We can explicitly see this in the experimental results presented.  

Since the polarization degree of freedom is used in the QI scheme, scattering of polarization state of photons will affect the value of $S$ affecting the detection of object.  The depolarizing noise on signal photons can be modelled using a path dependent depolarizing channel and the final state will be,

\begin{align}
D\left [ \rho (\eta) \right ]  = & \frac{p}{3} \left (\sum_{i =1}^{3} f_i  \rho (\eta) f_i^{\dagger}    \right )  + (1-p) \rho (\eta)
\end{align}
where  $f_i = \mathbbm{1} \otimes |0\rangle \langle 0 |  + \sigma_i \otimes |1\rangle \langle 1 |$ and $\sigma_i$ are the Pauli operators.  By subjecting  $D \left [ \rho(\eta)\right ]$ to the HWP and PBS as shown in Fig.\,\ref{QI-inter} for different values of $\theta$ and $\delta$ we can obtain CHSH parameter.  The theoretical expectation is presented in the following section along with the experimental results.


\section{Experimental Method }
\label{Exptsch}

	\subsection{Experimental Setup}
	The schematic of the experimental setup used for QI in this report using heralded single-photon entangled in polarization and path degree of freedom is shown in Fig.\,\ref{QI-inter}(b), non-interferometric approach. A 10-mm-long periodically-poled potassium titanyl phosphate (PPKTP) nonlinear crystal (Raicol) with poling period $\Lambda$ $=$ 10 \textmu m and aperture size of 1x2 mm$^{2}$ is deployed to generate heralded single photons using type-II SPDC process. The crystal is pumped using  continuous-wave diode laser (TopMode 405, Toptica) at 405 nm with 5 MHz linewidth. A half-wave plate is used to set the polarization of the laser and a plano-convex lens of 300 mm is used to focus the pump beam into the center of the crystal with beam waist \textit{w}\textsubscript{0} $=$ 42.5 \textmu m. PPKTP crystal is housed in an oven and its temperature is maintained at 23 \protect\raisebox{0.5pt}{$^\text{o}$}C to obtain degenerate photon pairs at 810 nm. We used a bandpass interference filter at 810 nm center wavelength with a bandwidth of 10 nm FWHM for collecting the SPDC photons from the residual pump light. The wavelength of the down-converted photons is confirmed using a spectrometer (QEPro, Ocean Insight). The generated orthogonally polarized photon pairs ($|h\rangle$, $|v\rangle$) are collimated using a plano-convex lens of 35 mm and separated using a polarization beam splitter.  The idler photons $|h\rangle$  which are used as reference for heralding are coupled to single-mode optical fiber using appropriate collection optics and sent directly to the receiving unit.  The signal photons in free space is passed through  a half-wave plate at $\pi/8$ and a polarizing beam splitter (PBS)  to generate a polarization-path entangled state. From the two pathways of the polarization-path entangled photons, the reference path is also coupled to single-mode fiber and sent to the receiving unit where as the signal path is sent towards the object. A non-polarizing beam splitter (BS) with variable reflectivity is used as an object in the signal path.  A broadband thermal light source is used to add noise into the system through another input port of the object BS and the photons from the signal path are collected at the receiving unit.  At the receiving unit,  HWP and PBS are placed along both, the signal and reference path.  The path length of the idler photon used for heralding is adjusted to the path length of the signal path by using the maximum coincidence counts for a fixed time window as reference point. Similar path length is set to reference path as well.  The output from the  both PBS are connected to four fiber coupled detectors, single photon counting modules, SPCM$_{j}$ (SPCM-800-44-FC, Excelitas) and the idler photons is also connected to another SPCM$_{5}$.   All the five detectors are connected to time-correlated single-photon counter (Time Tagger, Swabian instruments).  By taking the coincidence counts of photons from the four detectors with idler photon, probabilities of the basis states of the polarization-path composition of the photon are measured,

\begin{align}
P_{v1} (\theta, \delta)= \frac{ C_{1,5}(\theta, \delta) }{\sum_{j=1}^{5}C_{j,5} (\theta, \delta)} ~~;~~
P_{h1} (\theta, \delta)  = \frac{ C_{2,5}(\theta, \delta) }{\sum_{j=1}^{5}C_{j,5} (\theta, \delta)} \nonumber
  \end{align}
\begin{align}
P_{v0} (\theta, \delta)=
  \frac{ C_{3,5}(\theta, \delta) }{\sum_{j=1}^{5}C_{j,5} (\theta, \delta)} ~~;~~
  P_{h0} (\theta, \delta)  = \frac{ C_{4,5}(\theta, \delta) }{\sum_{j=1}^{5}C_{j,5} (\theta, \delta)}. 
\end{align}
$C_{j,5}(\theta, \delta)$ are the number of coincidence detection of photons in SPCM$_j$ and SPCM$_5$. Using these probabilities for different combination of $(\theta, \delta)$ we can calculate CHSH parameter $S$.  For the set of angles  $(\theta, \delta, \theta^{\prime}, \delta^{\prime}) = (0,  \pi/16, 3\pi/16,  5\pi/16 )$ realised using HWPs angles $(0,  \pi/32, 3\pi/32,  5\pi/32 )$ we get maximum S value. 

 \begin{figure}[h!]
\centering
\includegraphics[width=0.49\textwidth]{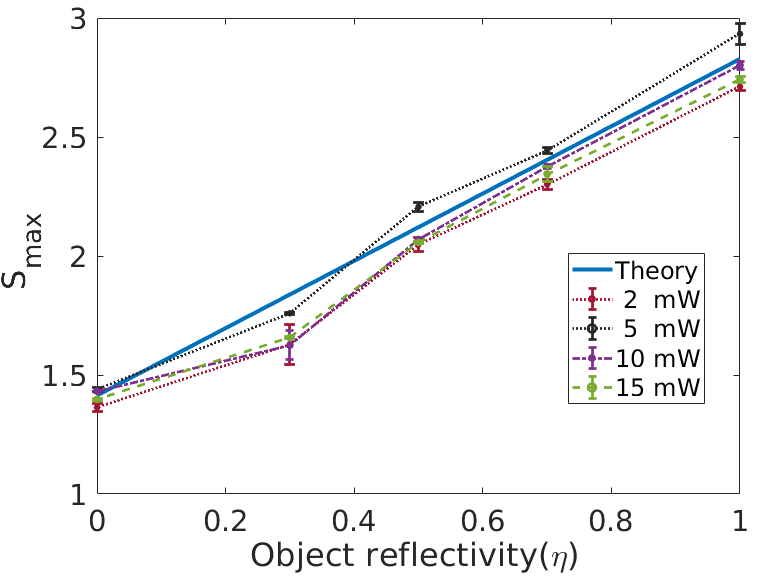}
\caption{Experimentally obtained maximum value of CHSH parameter $S_{\mbox{max}}$ for an object of different reflectivity.  $S_{\mbox{max}}$ was calculated using coincidence counts of four signal detectors with one heralding detector. The solid blue curve is the theoretical plot using a non-interference scheme. The red, black, violet and green data points are for a pump power of 2, 5, 10, and 15 mW, respectively. The error bars are for the standard deviation of the measurements. Experimental results obtained for different pump powers are in close agreement with the theoretical value.}
\label{expt-result}
\end{figure}

\subsection{Experimental result and analysis}

 \begin{figure}[htbp]
\centering
\includegraphics[width=0.49\textwidth]{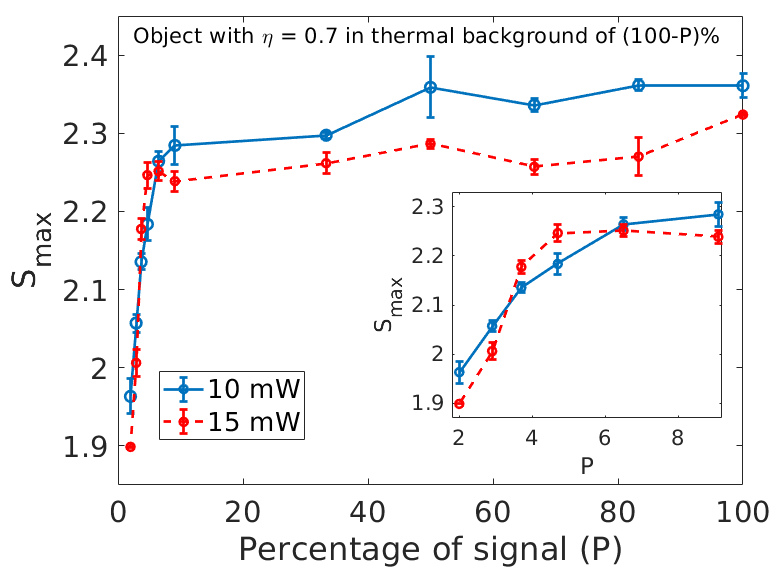}
\caption{ Experimentally obtained maximum value of CHSH parameter $S_{\mbox{max}}$ for different percentage of signal (P) in presence of an object of reflectivity $\eta = 0.7$. The background noise is increased such that the percentage of signal varied from 100 to 2. The solid and dashed line data points are for a pump power of 10 and 15 mW, respectively. The error bars are for the standard deviation of the measurements. The inset shows the zoomed region of the percentage of signal from 9 to 2.  We can see that $S_{\mbox{max}} > 2$  even when SNR=$0.03$.}
\label{expt-result1}
\end{figure}
 \begin{figure}
\centering
\includegraphics[width=0.49\textwidth]{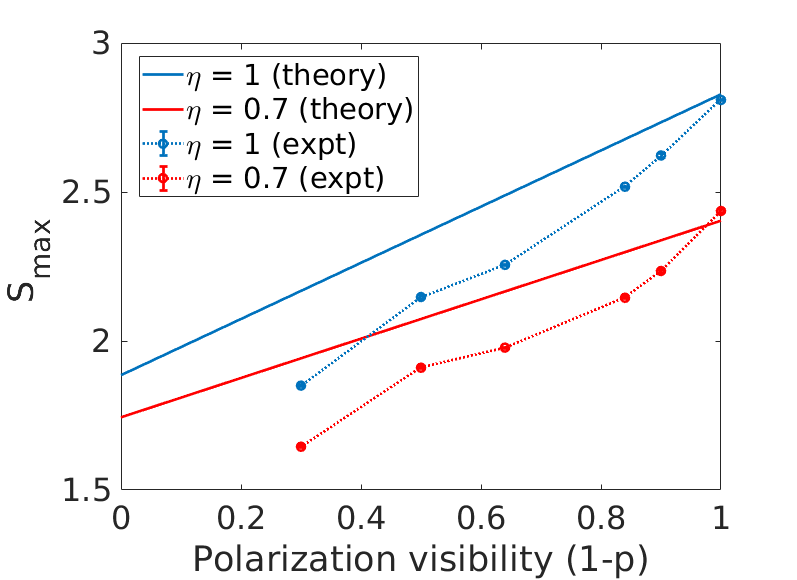}
\caption{Maximum value of CHSH parameter $S_{\mbox{max}}$ expected and experimentally obtained with change in polarization visibility. The results are for object reflectivity $\eta =1$ and $\eta =0.7$. We see a deviation of the experimental result from the theoretical value with decrease in visibility.} 
\label{expt-result2}
\end{figure}

In Fig.\,\ref{expt-result}, the maximum value of CHSH parameter, $S_{\mbox{max}}$ experimentally obtained when object of different reflectivity $\eta$ is illuminated using polarization-path entangled photon is shown for the  non-interferometric scheme.   The solid curve without any error bars is the theoretical plot using a non-interference scheme as shown in Fig.\,\ref{QI-interNon} for comparison. The red, black, violet, and green data points are for a pump power of 2, 5, 10, and 15 mW, respectively. The corresponding signal counts are $0.95 \times 10^5$,  $2.64 \times 10^5$, $4.45 \times 10^5$, and $6.93 \times 10^5$ counts/s, respectively. We can clearly note that the experimental value for different object reflectivity and for different pump power are all in close agreement  with the theoretical expectation. For an object with reflectivity $\eta =0.3$, the estimated $S_{\mbox{max}}  = 1.6 \pm 0.05$. Even though we don't see the violation of Bell's inequality here as presented in the theoretical description, for the value of $1.44  < S  < 2$  we can still infer the presence of object with low reflectivity.  

 Fig.\,\ref{expt-result1}  shows  $S_{\mbox{max}}$ for object reflectivity $\eta =  0.7$ when background noise was introduced.  Data points with the solid and dashed lines are for a pump power of 10 and 15 mW, respectively.  By keeping the number of signal photons fixed by fixning pump power, we increased the background noise such that the percentage of noise is varied from 0\% to 98 \% and calculated $S_{\mbox{max}}$.  When signal is only 10\% and noise is 90\%, SNR = 0.11.   One can see that with increasing background noise, the $S_{\mbox{max}}$  value remains almost same till signal is only 3\%, SNR$=0.07$  and with further increase in noise percentage it significantly reduces from $2.36\pm 0.04$ to $1.97\pm 0.04$.   Even when SNR = 0.02 which corresponds to -15 dB, we can note that  $S \approx 1.9$.  Using this result and the results reported in Fig.\,\ref{expt-result}, even when  $\eta = 0.2$  and SNR $=0.11$, $S_{\mbox{max}} > 1.5$  and hence can effectively be used as an indicator of presence of object with  very low reflectivity in the background noise. The object illuminated using polarization-path entangled photons can isolate the noise to a significant extent and register photons correlated only in photon correlated in polarization and path degree of freedom. We can clearly note from the inset figure that the value of $S > 2.2$ even when signal level is only 5\% of the background noise level entering the detector.  Only when background noise is $>95\%$ we see a significant and sudden dip in the value of $S$. That is the point beyond which photon number fluctuations from the  noise starts coinciding with the SPDC photons contributing to the change in photon counts due to the change in $(\theta, \delta)$ in the receiving unit. This makes polarization and path entangled QI scheme a highly robust even in high noise regime.

In Fig.\ref{expt-result2}, the maximum value of $S$ as a function of depolarization in the form of polarization visibility of photons from signal path when received from the object with $\eta = 1$ and $\eta=0.7$ is shown. Theoretical expectation is obtained using a depolarizing channel along the signal path.  For the experimental value, the polarization visibility is used to mimic depolarizing channel. It is realized by changing the combination of waveplates along the signal path. We can see that the experimental results obtained for polarization visibility in the range of 0.3  to 1 is lower than the theoretical value but they follow a similar trend. By collaborating the observations, even for the combination of depolarization effect, reflectivity and thermal noise the value of $S > 1.5$ can be used as an indicator of presence of object. 

The scheme will be highly effective when  all the three path lengths match. In real time scenario path lengths can be estimated by looking for the consistent match of the coincidence and single photon detection of the idler photons with the signal photons in schemes using SPDC process. When an on-demand single photon sources are used, time of arrival and time difference will help in ranging.





\section{conclusion}
\label{conc}
n summary, we demonstrate the use of polarization and path entangled single photons for QI that show that it detect low reflectivity object in noisy background where SNR is very low. Using heralded single photons from SPDC process we experimentally prepared the maximally entangled state in polarization and path degree of freedom of a single photon as the optimum probe state to illuminate the object.  We have reported the quantum advantage of using polarization-path entangled single photons over only single photons for QI.  The scheme will also provide a advantage over the fragility of two photon entangled state.  For an object reflectivity, $\eta > 0.5$ violation of Bell's inequality, $S \geq 2$ will confirm the presence of object even when the SNR is as low as 0.03. In addition to that, we have also shown the regime of classical correlation that are residual of quantum correlation to identifying object with $\eta < 0.5$. The non-interferometric measurement scheme we have demonstrated isolates the background noise from the signal and only signal photons contributes towards calculating $S$ value. Only when SNR $> 0.02$ we start seeing noise taking prominence. The results showing ability to detect object with reflectivity as low as $\eta =0.2$ in the background noise with SNR $=0.05$ demonstrate the robustness of the scheme. The scheme also suggests that the possibility of estimation of noisy environment by analyizing the photon measured and the deviation from the expected value at the receiving unit.  Further extension of this work using other internal degrees of freedom of photons may cover spectra of possibilities of using entangled single photon states for illumination, imaging and metrology tasks.   

\vskip 0.2in

{\bf Acknowledgment: } We thanks R. P. Singh and Somshubhro Bandyopadhyay for insightful comments and discussions.  We  also thank R. S. Gayatri for her support in laboratory during preparation stage of this experiment.  We acknowledge the financial support from the Office of Principal Scientific Advisor to Government of India, project no. Prn.SA/QSim/2020.


\end{document}